\documentclass[letterpaper]{article} 
\usepackage{aaai25}  
\usepackage{times}  
\usepackage{helvet}  
\usepackage{courier}  
\usepackage[hyphens]{url}  
\usepackage{graphicx} 
\usepackage{natbib}  
\usepackage{caption} 
\usepackage{algorithm}
\usepackage{algorithmic}
\usepackage{threeparttable}
\usepackage{booktabs}
\usepackage{amsopn}
\usepackage{multirow}
\usepackage{makecell}
\usepackage{tabularx}
\usepackage{bbding}
\usepackage{amsmath}
\usepackage{amssymb}
\usepackage{amsfonts}
\usepackage{bm}
\usepackage{array}
\usepackage{appendix}
\usepackage{adjustbox}
\usepackage{xspace}
\usepackage{arydshln}
\newcommand{\eg}{\textit{e.g.}\@\xspace}

\makeatletter
\DeclareRobustCommand\onedot{\futurelet\@let@token\@onedot}
\def\@onedot{\ifx\@let@token.\else.\null\fi\xspace}
\def\eg{\emph{e.g}\onedot}
\def\ie{\emph{i.e}\onedot}

\def\etal{\emph{et al}\onedot}
\makeatother

\usepackage{newfloat}
\usepackage{listings}
\DeclareCaptionStyle{ruled}{labelfont=normalfont,labelsep=colon,strut=off} 
\floatstyle{ruled}
\newfloat{listing}{tb}{lst}{}
\floatname{listing}{Listing}

\title{Patch-level Sounding Object Tracking for Audio-Visual Question Answering}

\author{
    Zhangbin Li\equalcontrib,
    Jinxing Zhou\equalcontrib,
    Jing Zhang,
    Shengeng Tang,
    Kun Li,
    Dan Guo\thanks{Corresponding authors.}
    \\
}
\affiliations{
    School of Computer Science and Information Engineering, Hefei University of Technology\\
    lizhangbin.mail@gmail.com, zhoujxhfut@gmail.com, guodan@hfut.edu.cn\\
}

\begin{document}

\maketitle

\begin{abstract}
Answering questions related to audio-visual scenes, \ie, the AVQA task, is becoming increasingly popular.
A critical challenge is accurately identifying and tracking sounding objects related to the question along the timeline.
In this paper, we present a new Patch-level Sounding Object Tracking (PSOT) method.
It begins with a Motion-driven Key Patch Tracking (M-KPT) module, which relies on visual motion information to identify salient visual patches with significant movements that are more likely to relate to sounding objects and questions.
We measure the patch-wise motion intensity map between neighboring video frames and utilize it to construct and guide a motion-driven graph network.
Meanwhile, we design a Sound-driven KPT (S-KPT) module to explicitly track sounding patches.
This module also involves a graph network, with the adjacency matrix directly regularized by audio-visual correspondence map.
The M-KPT and S-KPT modules are performed in parallel for each temporal segment, allowing balanced tracking of salient and sounding objects.
Based on the tracked patches, we further propose a Question-driven KPT (Q-KPT) module to retain patches highly relevant to the question, ensuring the model focuses on the most informative clues.
The audio-visual-question features are updated during the processing of these modules, which are then aggregated for final answer prediction.
Extensive experiments on standard datasets demonstrate the effectiveness of our method, achieving competitive performance even compared to recent large-scale pretraining-based approaches.
\end{abstract}

\section{Introduction}\label{sec:intro}

\begin{figure}[!t]
 \centering
 \includegraphics[width=1\columnwidth]{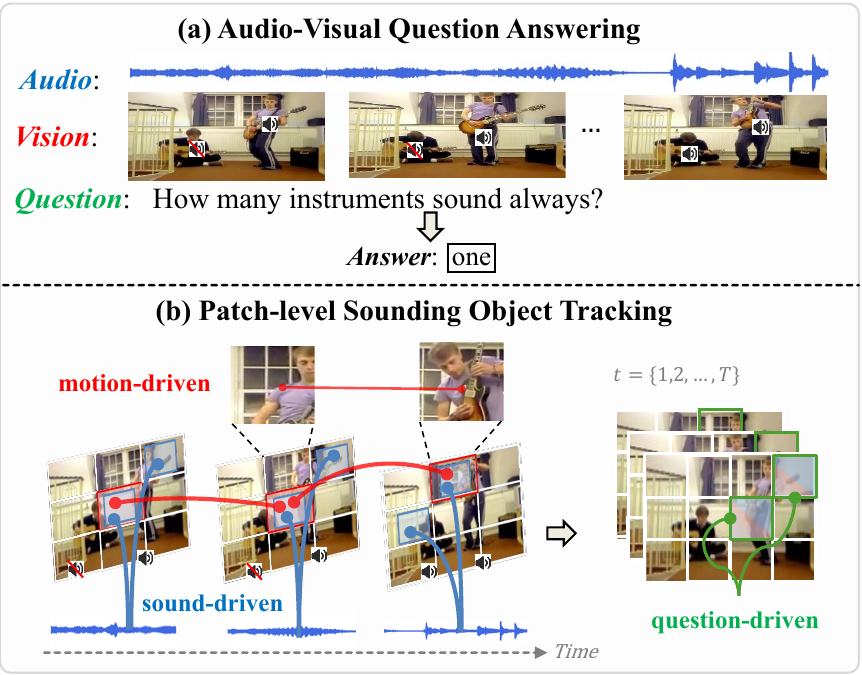}
    \caption{
    \textbf{Illustration of the AVQA task and our main idea.} (a) AVQA task requires accurate comprehension of sounding objects related to the question.
    (b) Our method explores the visual motion information and audio-visual correspondence to identify key salient and sounding patches. The question is then used to select the highly relevant patches.
    }
    \label{Fig:intro}
\end{figure}

Audio-Visual Question Answering (AVQA) is an attractive task in multimodal learning that involves the comprehension and utilization of audio, visual, and text modalities~\cite{mao2024tavgbench,shen2023fine, guo2024benchmarking, wang2024frequency}.
As shown in Fig.~\ref{Fig:intro}(a), 
given a long audible video and a related question, the AVQA task requires predicting the correct answer, necessitating the accurate perception of the pivotal audio-visual signals that are relevant to the question.
\textit{Mainstream} methods on the AVQA task focus on designing efficient expert models.
With limited training data, these methods emphasize the spatial-temporal reasoning of audio-visual scenes by exploring sounding regions~\cite{li2022learning}, question-related temporal segments~\cite{jiang2023target,li2023progressive}, or object-level cross-modal relations~\cite{li2024object}.
In particular, some methods attempt to utilize the frozen parameters of ViT~\cite{dosovitskiy2020image} models and insert learnable adapter layers to explore audio-visual relations, making the model suitable for the AVQA task~\cite{lin2023vision, duan2024cross}.
\textit{Another branch} of methods relies on large-scale pre-training.
For example, some works~\cite{shu2023audio, tang2024avicuna} attempt to extend large language models, \eg, Vicuna~\cite{touvron2023llama}, into multimodal scenarios, making them capable of downstream AVQA task.
Recently, VALOR~\cite{chen2023valor} introduced a pretraining model that generates unified and advanced audio-visual-language representations, which generalizes well to AVQA task.
However, all these methods require high computational resources or large-scale multimodal data during model pretraining/fine-tuning (we will provide more discussions and comparisons in the experiment section).
Our work belongs to the mainstream branch, \ie, building an expert AVQA model with sophisticated modules and a handful of training data.

Questions in the AVQA task usually involve some specific sounding objects.
As the example shown in Fig.~\ref{Fig:intro}(a), the question ``\textit{How many instruments sound always?}'' requires identifying the number of sounding instruments.
However, accurately identifying and perceiving the sounding objects occurring in the video is not easy.
\textbf{First}, the audio signal may be unclear, making it difficult to guide the identification of sounding objects.
Moreover, the audio signal may fail to distinguish objects with identical appearances (same categories).
For example, in Fig.~\ref{Fig:intro}(a), even if there is a clear sound of the guitar in the last segment, we cannot identify which guitar or whether both guitars emit the sound (instance-level identification).
\textbf{Second}, the sounding objects can be changeable along the timeline.
A sounding object may be salient in some temporal segments, while other salient objects may start to make sounds.
This requires modeling/tracking the audio-visual correspondences for each temporal segment.
\textbf{Third}, multiple objects may emit sounds. It is not enough to just perceive these sounding objects, but also to find the ones that are relevant to the given questions, \eg, the guitar on the right ``\textit{sound always}'' as queried in the question. 
According to these observations, we design meticulous modules and contribute a new \textbf{Patch-level Sounding Object Tracking (PSOT)} method for the AVQA problem.

\textbf{First}, to mitigate the potential deficiency of relying solely on audio signals, we propose to initially exploit the salient objects based on the visual information alone.  
Specifically, we identify salient visual patches with large movements, which are more likely to be sounding objects and helpful for question answering.
As shown in Fig.~\ref{Fig:intro}(b), the man on the right is playing the guitar, and the relevant visual patches significantly change between adjacent frames. 
In contrast, the ceiling and the floor are constantly static, providing useless clues.
To this end, we propose a \textbf{Motion-driven Key Patch Tracking (M-KPT)} module, which assesses the motion information between adjacent video frames by calculating the patch-wise feature similarity.
The generated motion intensity matrix is used to activate the salient visual patches.
Afterward, we construct a motion-driven graph network to propagate messages between features of visual patches and the synchronized audio, where the audiovisual nodes are guided by the motion intensity-driven adjacency relations.
Concretely, for each temporal segment, the audio and original visual patches are treated as the initial \textit{nodes} in the graph.
Then we initialize the adjacency matrix (\ie, \textit{edge set}) by computing feature similarity between audio and \textit{motion-enhanced} visual patches.
In this way, the visual patches of salient objects are enhanced and tracked through video segments, and the audio feature updating also benefits from the tracked salient objects.

\textbf{Meanwhile}, to explicitly model the audio-visual correspondence between audio and visual nodes, we design a \textbf{Sound-driven Key Patch Tracking (S-KPT)} module.
Similar to the M-KPT module, the S-KPT module involves a graph network.
These two modules share the same initial audiovisual nodes, and the primary difference lies in the adjacency matrix initialization.
In the S-KPT module, we directly measure the audio-visual correspondence map to enhance sounding patches.
Then, we generate the sound-driven adjacency matrix by computing feature similarity between audio and \textit{sound-enhanced} visual patches.
Notably, both audio and visual features are updated through the cross-modal interactions.
The M-KPT module and S-KPT module are deployed in parallel, allowing the model to dynamically integrate the visual motion guidance or audio guidance.

The above two modules can identify salient and sounding visual patches.
\textbf{Furthermore}, we design a \textbf{Question-driven Key Patch Tracking (Q-KPT)} module to identify and retain only those patches that are highly relevant to the question.
This is also implemented in a graph network.
We utilize the feature similarity (adjacency relations) between question and visual patches and retain the key patches with the highest pre-set percentage similarity values.
For those visual patches not retained, their features are filled with zeros and would not be learned during graph learning.
The Q-KPT ensures the model focuses more on the most informative visual patches related to the question. 
Given these preparations, we finally design a \textit{Multimodal Message Aggregation} module to aggregate information from enhanced audio-visual-question features for answer prediction.

In summary, our main contributions are fourfold:
\begin{itemize}
    \item By exploring patch-level clues of sounding objects, we present a novel graph-based PSOT model for the AVQA task. 
    Compared with large-scale pretraining-based methods, our model achieves competitive performance.
    \item We propose the M-KPT module to identify salient visual patches with large movements. To the best of our knowledge, we are also the first to explore patch-level motion information in the AVQA task.
    \item We propose the S-KPT module to identify sounding patches. The S-KPT cooperates with the M-KPT, guaranteeing effective audio-visual representation learning via visual motion and audio guidance.
    \item We propose the Q-KPT module to further identify patches highly related to the question, enabling the model to utilize more informative information for answering.
\end{itemize}

\begin{figure*}[t]
\begin{center}
\includegraphics[width=1\linewidth]{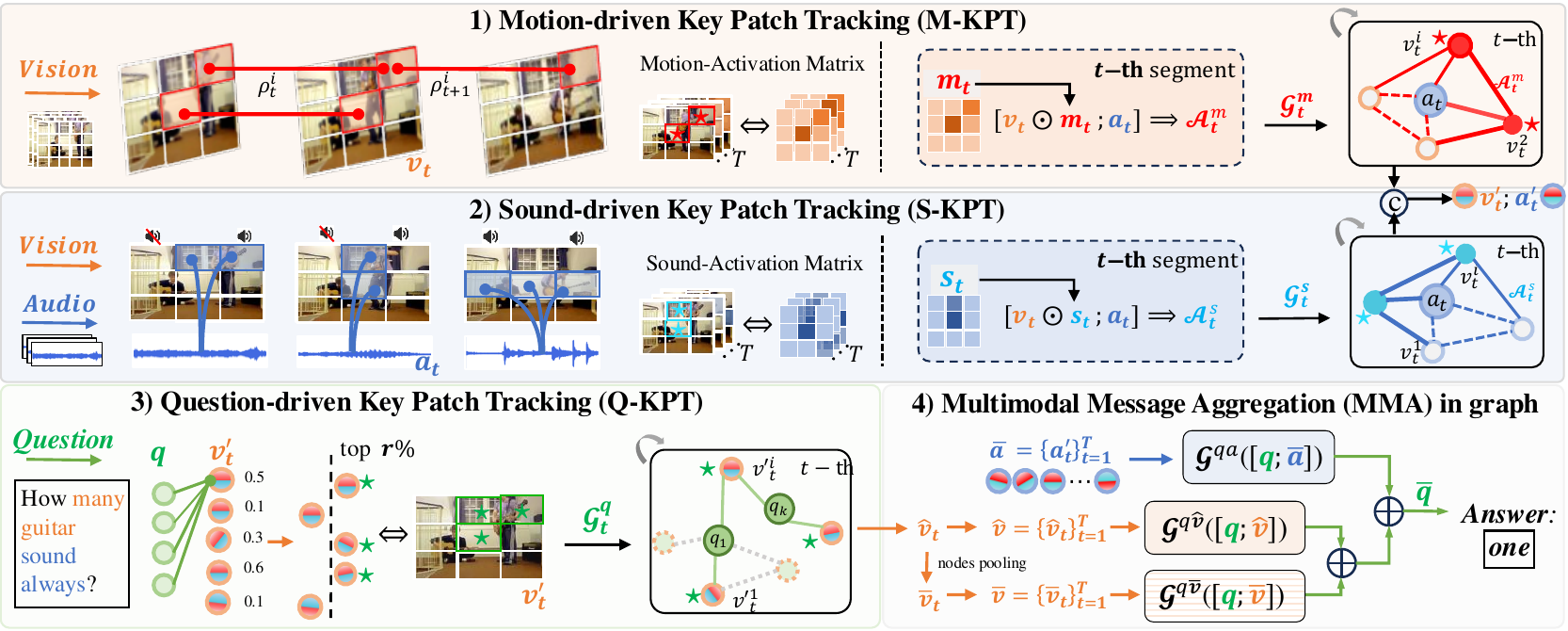}
\end{center}
\caption{
\textbf{Method Overview.}
1) The M-KPT module tracks salient visual patches ($\star$) with large motion, which often relate to sounding objects and questions. 
It measures the patch-wise motion intensity information between neighboring frames, yielding the motion-activation matrix ($\bm{m}_t$) to guide adjacency matrix ($\mathcal{A}^{m}_{t}$) for motion-driven graph network ($\mathcal{G}^m_t$) learning.
2) Meanwhile, the S-KPT module tracks the sounding patches by assessing the audio-visual correspondence. A sound-driven graph network ($\mathcal{G}^s_t$) is constructed.
3) The Q-KPT module further processes visual patches highlighted by the M-KPT and S-KPT modules, retaining only those patches highly relevant to the question. This is also achieved in a graph ($\mathcal{G}^q_t$).
4) The MMA module integrates the question with enhanced audio\&visual features through independent graph networks for answer prediction.
}
\label{fig:model}
\end{figure*}

\section{Related Work}

\noindent\textbf{Audio-Visual Question Answering.}
The AVQA task involves understanding audio-visual scenes~\cite{zhou2023contrastive,zhou2024label,zhou2024vaplan,zhou2024towards,zhou2022avs,zhou2023avss,zhou2021positive,zhou2023improving,guo2023audio} via a question-answering paradigm.
To address the AVQA task, earlier works~\cite{li2022learning,li2023progressive,duan2024cross} emphasize spatial-temporal reasoning by perceiving sounding visual regions and question-related temporal segments.
Recently, Li \etal~\cite{li2024object} utilize off-the-shelf object detection models to explicitly generate object-level visual features and then explore the relations among question, audio, and visual objects.
Meanwhile, other works address the AVQA task by exploring data bias~\cite{lao2023coca}, missing modalities~\cite{park2024enhancing}, and large-scale data pretraining techniques~\cite{shu2023audio,tang2024avicuna}.
We solve the AVQA problem by exploring patch-level tracking of motion-aware, sound-aware, and question-aware patches.

\noindent\textbf{Graph Neural Networks in QA tasks.}
GNNs are flexible in encoding intra- and cross-modal relationships and facilitating message passing among nodes, making them popular backbones in traditional QA tasks such as Visual Question Answering (VQA)~\cite{park2021bridge, de2023selfgraphvqa}, Video Question Answering~\cite{xiao2022video, xiao2023contrastive}, and text-based VQA~\cite{hu2020iterative, biten2022latr}.
In the AVQA field, mainstream methods typically employ Transformer~\cite{vaswani2017attention} blocks.
Recently, some works have explored using the Mamba~\cite{gu2023mamba} architecture to solve the AVQA problem~\cite{huangav,yang2024shmamba}.
We implement our approach with the exploration of GNN, demonstrating competitive performance.
Compared to traditional QA works, our graph-based method for AVQA has unique designs in constructing motion- and sound-driven adjacent matrices.

\section{Our Method}\label{sec:method}
In the AVQA task, an audible video is first divided into $T$ non-overlapping temporal segments. Each segment is represented by a 1-second audio signal and one visual frame.
Let us denote the extracted preliminary audio and visual features as $\bm{A}=\{\bm{a}_t\}_{t=1}^T \in \mathbb{R}^{T \times d}$ and $\bm{V} =\{\bm{v}_t\}_{t=1}^T \in \mathbb{R}^{T \times N^2 \times d}$, respectively.
Here, $d$ is the feature dimension, and $N^2$ is the number of patches (height $\times$ width) in each video frame.
As illustrated in Fig.~\ref{fig:model}, our framework for AVQA aims to identify the salient and sounding objects related to the question in each temporal segment and track them throughout the temporal sequences.
We detail each proposed module next.

\subsection{Motion-driven Key Patch Tracking}

As discussed in Introduction, the audio may be unclear and sometimes inefficient in identifying sounding objects sharing identical/similar appearances.
To alleviate this issue, we consider first relying on visual information to highlight salient visual patches with large movements, which are more likely to be related to sounding objects and the given question.
To this end, we propose a Motion-driven Key Patch Tracking (M-KPT) module.

Instead of using time-consuming optical flow~\cite{horn1981determining}, we calculate the discrepancy between adjacent frames at the patch-wise feature level to measure motion information.
Specifically, for the $i$-th patch of the $t$-th video frame, whose feature is denoted as $\bm{v}_t^i \in \mathbb{R}^{d}$, we compute its motion intensity $\rho_{t}^{i}$ as follows:  
\begin{equation}
    \rho_{t}^{i} = 1 - \frac{\bm{v}_t^{i} \cdot  \bm{v}_{t+1}^{i}}{\|\bm{v}_t^{i}\| \|\bm{v}_{t+1}^i\|},
    \label{eq:mot_s}
\end{equation}
where $\bm{v}_{t+1}$ is the feature of adjacent $t$+1 frame and $\rho_{t}^{i} \in [0,2]$.
If features of the $i$-th visual patch between adjacent frames have low cosine similarity, it will have a high $\rho_{t}^{i}$, indicating that the $i$-th patch may be part of a salient object.
By performing Eq.~\ref{eq:mot_s} for each patch, we obtain the patch-wise motion intensity matrix for $t$-th frame, denoted as $\bm{\rho}_t$. 

$\bm{\rho}_t = \{\rho_t^i\}_{i=1}^{N^2} \in \mathbb{R}^{N^2}$ reflects the motion intensity by capturing the patch-wise movement between adjacent video frames.
Furthermore, in some scenarios, some salient or informative visual patches may only have large movements in some temporal frames.
However, we believe that each video frame should emphasize these patches at the positions where the large movements exist in other frames.
To track objects with large movement along the timeline, we additionally consider the global motion intensity across all temporal frames, computed as: $\bm{\overline{\mu}} = \frac{1}{T} {\textstyle  \sum_{t=1}^T} \bm{\rho}_t$.
The total motion intensity matrix of $t$-th frame is a combination of the local motion matrix $ \bm{\rho}_t$ and the global matrix $\bm{\overline{\mu}}$: 
\begin{equation}
     \bm{m}_t = (1- \lambda) \bm{\rho}_t + \lambda  \bm{\overline{\mu}},
     \label{eq:motion_act_matrix}
\end{equation}
where $\bm{m}_t \in \mathbb{R}^{N^2}$ and $\lambda$ is a constant, empirically set to 0.2.

After obtaining the motion intensity matrix $\bm{m}_t$ for the $t$-th frame, we highlight those salient visual patches within the frame by multiplying $\bm{m}_t$ with $\bm{v}_t$: $ \bm{v}_t^m = \bm{m}_t \odot \bm{v}_t$, where $\odot$ is the Hadamard product.
Then, the motion-enhanced visual patch features $\bm{v}_t^m \in \mathbb{R}^{N^2 \times d}$ are concatenated with the audio feature $\bm{a}_t \in \mathbb{R}^{1 \times d}$, which is then used to calculate the initial motion-driven adjacency matrix $\mathcal{A}_t^m \in \mathbb{R}^{(N^2+1) \times (N^2+1)}$. 
This process can be formulated as follows:
\begin{equation}
    \mathcal{A}_t^m  = [\bm{m}_t \odot \bm{v}_t; \bm{a}_t] [\bm{m}_t \odot \bm{v}_t;\bm{a}_t]^{\top},
    \label{Eq:adj_mot}
\end{equation}
where $[\cdot\, ;\cdot]$ represents the concatenation operator and $\mathcal{A}_t^m$ reflects the pairwise relations.

The original visual features $\bm{v}_t \in \mathbb{R}^{N^2 \times d}$ are concatenated with the audio feature $\bm{a}_t \in \mathbb{R}^{1 \times d}$ as the basic \textit{nodes} in the graph for $t$-th segment: $\mathcal{N}_t = [\bm{v}_t; \bm{a}_t] \in \mathbb{R}^{(N^2+1) \times d}$.
With the nodes $\mathcal{N}_t$ and adjacency matrix $\mathcal{A}_t^m$ in hand, we construct a motion-driven graph neural network $\mathcal{G}^m_t$ to encode the interactions between visual patches and audio features: 
\begin{equation}
        \mathcal{G}^m_t(\mathcal{N}_t) := \text{ReLU}(\mathcal{A}_t^m \mathcal{N}_t \bm{W}^m),
\label{Eq:standard}
\end{equation}
where $\bm{W}^m \in \mathbb{R}^{d \times d}$ is a learnable parameter. $\mathcal{N}_t$ is updated through $L$ iterations, denoting $L$ graph layers. We denote the audiovisual nodes after completing graph learning as $\mathcal{N}_t^L$.
Notably, we construct the graph $\mathcal{G}^m_t$ for each temporal segment, \ie, $t=1,2,...,T$. The parameters are shared among different segments, ensuring computation efficiency. 
In this way, the salient visual patches are tracked along the temporal sequence, highlighted by the motion intensity matrix, and are then used to interact with corresponding audio node, benefiting effective audio-visual representation learning.

\subsection{Sound-driven Key Patch Tracking}
The aforementioned M-KPT module relies on visual motion guidance and implicitly affects the audio node by interacting with motion-enhanced visual patches.
To explicitly model the audio-visual correlations, \ie, perceiving sounding patches, we propose the Sound-driven Key Patch Tracking (S-KPT) module.

Specifically, for the $t$-th segment, we directly assess the audio-visual correspondence between the audio node $\bm{a}_t \in \mathbb{R}^{1 \times d}$ and visual patch nodes $\bm{v}_t \in \mathbb{R}^{N^2 \times d}$. The sound-activation matrix $\bm{s}_t$ can be generated as follows:
\begin{equation}
    \bm{s}_t = \frac{ \bm{a}_t }{ \| \bm{a}_t \|} \otimes (\frac{\bm{v}_t}{\| \bm{v}_t \|})^{\top},
    \label{eq:sound_act_matrix}
\end{equation}
where $\otimes$ is the matrix multiplication and $\bm{s}_t \in \mathbb{R}^{N^2}$.

Similar to the M-KPT module, we further utilize the obtained $\bm{s}_t$ to identify sounding patches within the frame by multiplying $\bm{s}_t$ with $\bm{v}_t$.
Then, the sound-enhanced features of visual patches are concatenated with the audio feature $\bm{a}_t$ to compute the initial sound-driven adjacency matrix $\mathcal{A}_t^s \in \mathbb{R}^{(N^2+1) \times (N^2+1)}$:
\begin{equation}
    \mathcal{A}_t^s = [\bm{s}_t \odot \bm{v}_t; \bm{a}_t] [\bm{s}_t \odot \bm{v}_t;\bm{a}_t]^{\top}.
    \label{Eq:adj_aud}
\end{equation}

Given the audiovisual nodes $\mathcal{N}_t=[\bm{v}_t; \bm{a}_t]$, the sound-driven graph neural network can be formulated as,
\begin{equation}
    \mathcal{G}^s_t(\mathcal{N}_t) := \text{ReLU}(\mathcal{A}_t^s \mathcal{N}_t \bm{W}^s),
\end{equation}
where $\bm{W}^s \in \mathbb{R}^{d \times d}$ is the learnable parameter.
We perform $\mathcal{G}^s_t$ for each temporal segment for the sounding patch tracking, and the parameters are shared across segments.

As shown in Fig.~\ref{fig:model}, we perform the M-KPT and S-KPT modules in parallel, enabling a joint and balanced utilization of the inherent visual motion and audio information to track salient and sounding patches within temporal segments.
The updated audio-visual nodes output from the final layer of $\mathcal{G}_t^m$ and $\mathcal{G}_t^s$ are concatenated and sent to a fully connected layer for fusion.
We denote the obtained audio and visual nodes as $\bm{a}'_t \in \mathbb{R}^{1 \times d}$ and $\bm{v}'_t \in \mathbb{R}^{N^2 \times d}$ ($t=1,2,...,T$), respectively.

\subsection{Question-driven Key Patch Tracking}

After highlighting the salient and sounding patches, we further consider that some of them may be redundant for answering the flexible questions.
Therefore, we propose a Question-driven Key Patch Tracking (Q-KPT) module to retain only those patches within each frame highly relevant to the question, serving as refined clues for answering.

Given the word-level feature of the question $\bm{q} \in \mathbb{R}^{K \times d}$ ($K$ is the length of words) and the visual patches at $t$-th segment $\bm{v}'_t \in \mathbb{R}^{N^2 \times d}$, we first calculate their feature similarity ($\mathbb{R}^{K \times N^2}$), similar to Eq.~\ref{eq:sound_act_matrix}.
Then, an average pooling operation along the word dimension is employed to obtain the sentence-level question-to-visual similarity, denoted as $\bm{\alpha}_t \in \mathbb{R}^{N^2}$.
Then, we propose to only reserve those visual patches that are among the top $r$(\%) in terms of relevance to the question, as reflected in $\bm{\alpha}_t$, for each frame.
The mask matrix $\bm{\beta}_t \in \mathbb{R}^{N^2}$ indicating whether the $i$-th patch is retained can be computed as,
\begin{equation}
    {\beta}_t^i =
    \left\{\begin{matrix}
        \begin{aligned}
                1 \;, \; {\alpha}_t^i \in \text{top}\; \; r \; \; \text{of} \; \;  \bm{\alpha}_t; \\
                0 \;, \; {\alpha}_t^i \notin \text{top}\; \; r \; \; \text{of} \; \; \bm{\alpha}_t.
        \end{aligned}
    \end{matrix}\right.
    \label{Eq:sparse}
\end{equation}
In this way, only $N^2*r$ visual patches are retained in each frame.
For other patches (where $\beta_t^i = 0$), their features are set to zeros in our implementation and will not be learned.

In fact, the values of $\bm{\alpha}_t$ and $\bm{\beta}_t$ can be computed and dynamically learned in a graph network.
Specifically, the words of the question $\bm{q} \in \mathbb{R}^{K \times d}$ and the visual patches $\bm{v}'_t \in \mathbb{R}^{N^2 \times d}$ are concatenated as the basic nodes $\mathcal{M}_t$ in a graph.
We can calculate the initial adjacency matrix $\mathcal{A}_t^q \in \mathbb{R}^{(K+N^2)\times(K+N^2)}$ as follows: 
\begin{equation}
    \mathcal{A}_t^q = \mathcal{M}_t \mathcal{M}_t^{\top} = [\bm{v}'_t; \bm{q}] [\bm{v}'_t;\bm{q}]^{\top}.
    \label{Eq:adj_qkpi}
\end{equation}
The $\bm{\alpha}_t$ can be obtained from $\mathcal{A}_t^q$ by utilizing only the adjacent weights between nodes of question words and visual patches.
Accordingly, we can construct the graph neural network $\mathcal{G}_t^q$, formulated as,
\begin{equation}
    \mathcal{G}^q_t(\mathcal{M}_t) := \text{ReLU}(\mathcal{A}_t^q \mathcal{M}_t \bm{W}^q),
\end{equation}
where $\bm{W}^q \in \mathbb{R}^{d \times d}$ is the learnable parameter.
For convenience, we denote the updated features of visual patches at $t$-th segment as $\bm{\hat{v}}_t \in \mathbb{R}^{N^2 \times d}$.
Notably, some visual patches have been discarded by setting their features to zeros.
Similar to $\mathcal{G}_t^m$ and $\mathcal{G}_t^s$, $\mathcal{G}_t^q$ is performed for each video frame for tracking the highly question-relevant patches along the timeline, and the parameters are shared across segments.

\subsection{Multimodal Message Aggregation}
After going through the above introduced three modules, we obtain the enhanced audio and patch-level visual features, denoted as, $\bm{\overline{a}} = \{ \bm{a}'_t \}_{t=1}^T \in \mathbb{R}^{T \times d}$ and $\bm{\hat{v}} = \{\bm{\hat{v}}_t \}_{t=1}^T \in \mathbb{R}^{T \times N^2 \times d}$, respectively.
For visual modality, the segment-level feature $\bm{\overline{v}} \in \mathbb{R}^{T \times d}$ can be obtained from $\bm{\hat{v}}$ via average pooling operation on the spatial dimension.

Given these outcomes, we propose the Multimodal Message Aggregation (MMA) module to comprehensively utilize the useful clues related to the question $\bm{q}$ for final answer prediction.
Here, we continue to adopt the graph network to model the interactions between audio/visual and question modalities.
Specifically, the question feature $\bm{q}$ is concatenated with the audio feature $\bm{\overline{a}}$, patch-level visual feature $\bm{\hat{v}}$, and segment-level visual feature $\bm{\overline{v}}$, respectively.
Accordingly, we compute their feature similarity as the initial adjacency matrix.
For simplicity, we abbreviate the constructed graph networks as $\mathcal{G}^{qa}([\bm{q}; \bm{\overline{a}}])$, $\mathcal{G}^{q{\hat{v}}}([\bm{q}; \bm{\hat{v}}])$, and $\mathcal{G}^{q\overline{v}}([\bm{q}; \bm{\overline{v}}])$, respectively.
The question nodes are updated in each graph network.
We then aggregate the learned features of the question nodes from the three graphs into a hidden feature $\bm{\overline{q}} \in \mathbb{R}^{L \times d}$, which is sent to a fully connected layer followed by a softmax function to generate the final probability $\bm{p}\in\mathbb{R}^C$ for the candidate answer predictions. The cross-entropy loss is used for model training.

\section{Experiments}
\subsection{Experimental Setup}
\textbf{Dataset and evaluation metric.}
We primarily conduct experiments on the widely-used and challenging MUSIC-AVQA~\cite{li2022learning} dataset.
This dataset consists of 9,288 videos involving various audio-visual music scenarios.
Using 33 predefined QA templates, over 45,000 question-answer pairs are constructed.
Most of the QA pairs are related to both audio and visual modalities (\textbf{AV-QA}), while the rest focus on either audio or visual modality alone (\textbf{A-QA}/\textbf{V-QA}).
Following the standard protocol in the pioneering work~\cite{li2022learning}, we adopt the answer prediction accuracy (\%) as the metric for model evaluation.

\noindent\textbf{Implementation details.}
Each video in the dataset is 60 seconds long.
To reduce computational costs, we sample one video frame every 6 seconds,
resulting in $T$=$10$ temporal segments.
Accordingly, we obtain 10 1-second audio segments synchronized with the video frames.
We employ the image encoder of pretrained CLIP-ViT-L/14~\cite{radford2021learning} model to extract 16$\times$16$\times$1024-D patch-level visual features, which are further downsampled to 8$\times$8$\times$1024 using average pooling to further reduce computational costs;
For audio features, we use CLAP~\cite{wu2023large} to extract 512-D representations;
For question embeddings, we utilize the text encoder of CLIP-ViT-L/14 to obtain 768-D word-level embeddings.
We then apply linear projections to convert these features to the same 512-D.
During model training, we use the AdamW optimizer with an initial learning rate of 1\textit{e}-4, which decays by 0.1 every 16 epochs.
The batch size and epochs are set to 16 and 35, respectively.
The numbers of graph layers in $\mathcal{G}_t^m$, $\mathcal{G}_t^s$, and $\mathcal{G}_t^q$ are empirically set to 3, 3, and 2, respectively.
All experiments are conducted on an NVIDIA A40 GPU.

\begin{table*}[!tbp]
  \scriptsize
  \centering
  \setlength\tabcolsep{12pt}
  \small
  {
   \begin{threeparttable}
   \resizebox{0.99\textwidth}{!}{
       \begin{tabular}{c|c|c|c|c|c|c}
            \Xhline{1.2pt}
            Method & Visual Encoder &Audio Encoder & \ \ A-QA\ \  & \ \ V-QA\ \  & \ \ AV-QA \ \  & \ \ Avg. \ \ \\
            \hline
            AVSD~\cite{schwartz2019simple} &VGG-19 &VGGish &68.78  &70.31 &65.44 & 67.32 \\
            PanoAVQA~\cite{yun2021pano} &Faster R-CNN &VGGish &72.13  &73.16 &66.97 &69.53 \\ 
            
            AVST~\cite{li2022learning} &ResNet-18 &VGGish &73.87	&74.40 &69.53	&71.59 \\
            
            COCA~\cite{lao2023coca} &ResNet-18 &VGGish &{75.42} &{75.23}	&69.96	&{72.33} \\
            
            TJSTG~\cite{jiang2023target} &ResNet-18 &VGGish &76.47 &76.88 &70.13 &73.04 \\

            AV-Mamba~\cite{huangav} &ResNet-18 &VGGish &74.67 &79.27 &70.62 &73.63 \\
            
            SH-Mamba~\cite{yang2024shmamba} &ResNet-18 &VGGish &75.42 &79.93 &70.64 &74.12 \\
            \hline
            
            APL~\cite{li2024object} &DETR &VGGish &78.09 &79.69 &70.96	&74.53 \\

            PSTP~\cite{li2023progressive}& CLIP$_\textbf{B}$ &VGGish &70.91	&77.26 &72.57	&73.52 \\

            LAVISH~\cite{lin2023vision} &Swin$_\textbf{L}$ & CLAP &75.40 &{79.60} &70.10 &73.60  \\

            CMP~\cite{duan2024cross} &Swin$_\textbf{L}$ & CLAP &77.40 &\textbf{81.90} &70.70 &74.80 \\ \hline
            
            \textbf{PSOT$_{\text{B}}$ (ours)} &CLIP$_\textbf{B}$ &VGGish &\textbf{78.22}	&80.07	&\textbf{72.61}	&\textbf{75.29} \\
            
            \textbf{PSOT$_{\text{L}}$ (ours)} &CLIP$_\textbf{L}$ & CLAP &\textbf{79.08} &\textbf{87.12} &\textbf{74.07} &\textbf{78.42} \\
           \Xhline{1.2pt}
       \end{tabular}}
        
    \end{threeparttable}}
    \caption{
        \textbf{Comparison with mainstream methods.} 
        The subscript `B' in `PSOT$_\text{B}$ and `L' in `PSOT$_\text{L}$ denote that the CLIP-ViT-B or CLIP-ViT-L model is used for visual feature extraction.
        The top-2 results are highlighted in \textbf{bold}. }
    \label{tab:music_avqa}
\end{table*}

\subsection{Comparison with Prior Works}
\textbf{Comparison with mainstream methods.}
Our method belongs to the mainstream branch of AVQA works, focusing on designing expert models with limited training data.
In Table~\ref{tab:music_avqa}, we compare our method with existing mainstream methods.
\textbf{1)} Early works on the AVQA task typically utilize basic ResNet-18 to extract visual features. Recent works have started employing more powerful visual encoders, such as CLIP$_\text{B}$ (CLIP-ViT-B/32), and Swin$_\text{L}$ (SwinT-V2-L~\cite{liu2021swin}).
Our method PSOT$_\text{B}$ employs CLIP$_\text{B}$/VGGish as the visual/audio feature extractor, the same as PSTP~\cite{li2023progressive}, and surpasses PSTP by 1.77\% in average QA accuracy.
Notably, our method also sightly outperforms the previous state-of-the-art method CMP~\cite{duan2024cross}, which utilizes more advanced visual and audio encoders.
These results demonstrate the effectiveness and superiority of our model design.
Our model's performance is significantly boosted when adopting more powerful CLIP$_\text{L}$ and CLAP models for feature extraction.
The average performance reaches 78.42\%.
Our subsequent experiments are based on this configuration.
\textbf{2)} Most mainstream works are implemented using Transformer~\cite{vaswani2017attention}.
Some recent methods exploit the Mamba architectures and achieve competitive performance, \eg, 74.12\% accuracy achieved by SH-Mambda~\cite{yang2024shmamba}.
Compared to these methods, our graph-based method demonstrates superior performance, providing new evidence of the potential of exploiting graph networks for the AVQA task.

\begin{table}[!tb]
  \scriptsize
  \centering
  \setlength\tabcolsep{1pt}
  \small
  {
   \begin{threeparttable}
   \resizebox{0.99\columnwidth}{!}{
       \begin{tabular}{c|c|c|c|c|c}
            \Xhline{1.2pt}
            Method & V- Enc. & A- Enc. & \#PT & Params & Acc. \\
            \hline
            AVLLM~\cite{shu2023audio} &CLIP$_\textbf{L}$  &CLAP &1.6M & 13B &45.20 \\
            AVicuna~\cite{tang2024avicuna} &CLIP$_\textbf{L}$ &CLAP &1.1M & 7B &49.60 \\ \hline
            VALOR~\cite{chen2023valor} &CLIP$_\textbf{L}$ &AST &33.5M & 593M &78.90 \\
            \hline
            \textbf{PSOT$_\textbf{L}$ (ours) } &CLIP$_\textbf{L}$ &CLAP &9k & 8.3M &78.42 \\
           \Xhline{1.2pt}
       \end{tabular}}
    \end{threeparttable}}
    \caption{
        \textbf{Comparison with large-scale pretraining-based methods.} `\#PT' denotes the number of samples used for model pretraining. `Params' is the model size of LLMs (for the first two methods) or the trainable parameters of the entire model (for the last two methods).}
    \label{tab:music_avqa_llm}
\end{table}

\begin{table}[!tb]
  \scriptsize
  \centering
  \setlength\tabcolsep{4pt}
  \small
  {
   \begin{threeparttable}
   \resizebox{0.99\columnwidth}{!}{
        \begin{tabular}{c | ccc|ccc|c}
            \Xhline{1.2pt}
            Id &M-KPT &S-KPT &Q-KPT & A-QA& V-QA & AV-QA& Avg.\\
            \hline
            a &$\checkmark$ &$\checkmark$ &$\checkmark$ &{79.08} &{87.12} &{74.07} &{78.42} \\ \hline
            b &-- &$\checkmark$ &$\checkmark$ &78.09  &87.12  &73.01  &77.65 \\
            c &$\checkmark$ &-- &$\checkmark$ &77.53  &86.99  &73.50  &77.79 \\
            d &$\checkmark$ &$\checkmark$ &-- &77.84  &86.21  &73.38  &77.57 \\ \hline
            e &-- &-- &$\checkmark$ &76.60  &85.34  &73.04  &77.05 \\
            f &-- &-- &-- &76.03 &83.45 &72.40 &75.88 \\
            \hline
            \Xhline{1.2pt}
        \end{tabular}
        }
    \end{threeparttable}}
    \caption{
        \textbf{Ablation study on the proposed main modules.}
    }
\label{tab:main_module_abltion} 
\end{table}

\noindent\textbf{Comparison with large-scale pretraining-based methods.}
The recent progress in the AVQA field tends to utilize either Large-Language Models (LLM) or large-scale data pretraining.
\textbf{1)} AVLLM~\cite{shu2023audio} and AVicuna~\cite{tang2024avicuna} are two representative LLM-based methods that employ the Vicuna-13B and Vicuna-7B~\cite{touvron2023llama}, respectively.
A common initial step in these methods is to align audio and visual features with the LLM's token embedding space through linear projections.
To achieve this, large-scale audio-caption and visual-caption pairs are used.  
Additionally, to adapt the LLM for processing audio and visual inputs and encoding cross-modal relations, these methods must fine-tune the LLM, which necessitates additional high-quality audio-visual-text data. 
As shown in Table~\ref{tab:music_avqa_llm}, AVicuna requires 1.1M data in total.
Since these methods do not fine-tune on the downstream MUSIC-AVQA dataset, their zero-shot performances are currently far behind those of supervised learning-based methods shown in Table~\ref{tab:music_avqa}.
\textbf{2)} The recent method VALOR~\cite{chen2023valor} focuses on audio-visual-language representation learning and utilizes 33.5M multimodal pairs for model pretraining.
In this way, VALOR can generate advanced tri-modal features and generalize well to the AVQA task.
After further fine-tuning on the MUSIC-AVQA dataset, VALOR achieves an impressive accuracy of 78.90\%.
In contrast, as shown in the Table, our method can achieve competitive performance while utilizing significantly less training data (9k vs. 33.5M) and fewer trainable parameters (8.3M vs. 593M).
Notably, VALOR employs visual features (16$\times$16$\times$1024) extracted by \textit{fine-tuned} CLIP$_\text{L}$ model while we use the \textit{downsampled} features (8$\times$8$\times$1024) from \textit{pretrained} CLIP$_\text{L}$.
These results validate the effectiveness of the proposed method, indicating that it is feasible to address specific downstream tasks by developing a resource-efficient expert model.

\subsection{Ablation Studies}
\textbf{Ablation study on main modules.}
We conduct ablation experiments to evaluate the effectiveness of the proposed main modules, \ie, the M-KPT, S-KPT, and Q-KPT.
The results are shown in Table.~\ref{tab:main_module_abltion}.
Compared to the complete model (a), the average QA performance after removing the M-KPT, S-KPT, or Q-KPT module (b/c/d) decreases by 0.77\%, 0.63\%, or 0.85\%, respectively.
When the M-KPT and S-KPT modules are simultaneously removed (e), the model performance further decreases, indicating that these modules performing in parallel complement each other.
The model's performance continues to deteriorate if all three modules are removed (f), \ie, the audio and visual features extracted by backbones are directly sent to the final MMA module for answer prediction.
These results clearly demonstrate the effectiveness of each proposed module.

\begin{table}[!t]
  \scriptsize
  \centering
  \setlength\tabcolsep{7pt}
  \small
  {
   \begin{threeparttable}
   \resizebox{0.99\columnwidth}{!}{
        \begin{tabular}{c | cc|ccc|c}
            \Xhline{1.2pt}

            Id & $\mathcal{A}_t^m$ & $\mathcal{A}_t^s$ & {A-QA} & {V-QA} & {AV-QA}  & {Avg.} \\
            \hline
            a &$\checkmark$ &$\checkmark$ &79.08 &{87.12}  &{74.07} &{78.42} \\
            \hline
            b &-- &$\checkmark$ &{79.14}  &86.75  &73.52  &77.92 \\
            c &$\checkmark$ &-- &79.02  &86.87  &73.66  &78.08 \\  
            \hline
            d  &-- &-- &77.59 &86.00 &73.01 &77.27 \\
            
            \Xhline{1.2pt}
        \end{tabular}
        }
    \end{threeparttable}}
    \caption{
        \textbf{Impacts of the motion-/sound-driven adjacency matrices} in M-KPT and S-KPT modules, \ie, $\mathcal{A}_t^m$ and $\mathcal{A}_t^s$ .
    }
\label{tab:ablation_ms_adj}
\end{table}

\begin{figure*}[!htb]
\begin{center}
\includegraphics[width=1\linewidth]{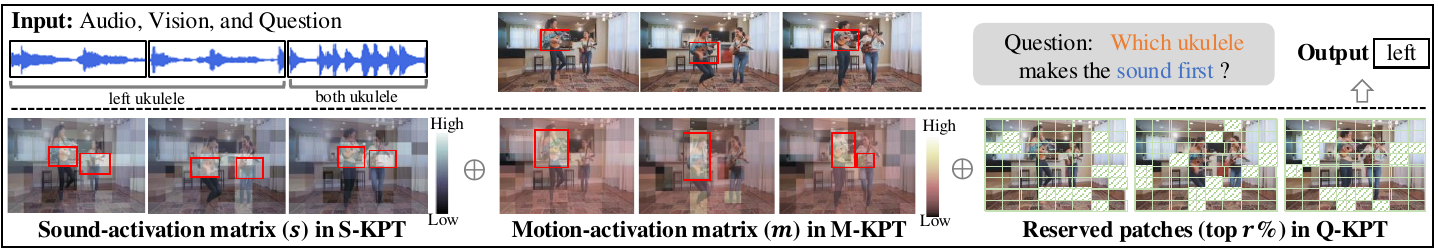}
\end{center}
\caption{
\textbf{Visualization of our model's inference process for AVQA.} The key patches (red boxes) are highlighted in each module, guided by sound, motion, and question. The brighter color indicates larger activation weights.
}
\label{fig:visual_flow}
\end{figure*}

\begin{figure*}[!htb]
\begin{center}
\includegraphics[width=1\linewidth]{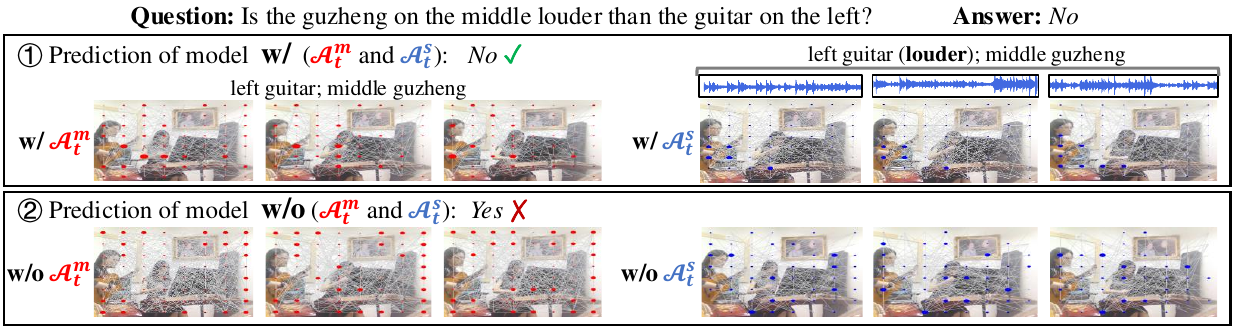}
\end{center}
\caption{
    \textbf{Ablation visualization on the roles of $\mathcal{A}_t^m$ and $\mathcal{A}_t^s$ in our model.} 
    The size of dots ($\bullet$) located in the center of visual patches represents the aggregated weight from the adjacency matrix during the graph network learning.
}
\label{fig:ablation_visual}
\end{figure*}

\noindent\textbf{Impacts of $\mathcal{A}_t^m$ and $\mathcal{A}_t^s$ in M-KPT and S-KPT modules.}
The $\mathcal{A}_t^m$ (Eq. \ref{Eq:adj_mot}) and $\mathcal{A}_t^s$ (Eq. \ref{Eq:adj_aud}) are the motion- and sound-driven adjacency matrices utilized in our M-KPT and S-KPT modules, respectively.
To examine their impacts, we compare them with a vanilla strategy that directly calculates the feature similarity between original concatenated audiovisual nodes as the adjacency matrix, \ie, $\mathcal{A}_t=[\bm{v}_t; \bm{a}_t] [\bm{v}_t;\bm{a}_t]^{\top}$.
As shown in Table~\ref{tab:ablation_ms_adj}, compared to the full model (a), the average model performance without utilizing $\mathcal{A}_t^m$ (b), $\mathcal{A}_t^s$ (c), or both (d) all decreases.
This proves that the motion intensity and audio-visual correspondence maps used for $\mathcal{A}_t^m$ and $\mathcal{A}_t^s$ computation are more useful for guiding graph learning.

\subsection{Qualitative Results}

\textbf{Visualization of our model's inference process.}
We provide a qualitative example to illustrate how our model works for audio-visual question answering. 
As shown in Fig.~\ref{fig:visual_flow}, the person on the left continuously plays the ukulele, while the person on the right holds a ukulele for a time before playing it in the final segment.
To answer the question ``\textit{Which ukulele makes the sound first?}'', our S-KPT module tracks the two instruments corresponding to the sound. As verified by the visualized sound-activation matrix $\bm{s}$ (Eq.~\ref{eq:sound_act_matrix}), the visual patches of the two ukuleles exhibit larger weights (brighter colors).
Since both ukuleles semantically correspond to the sound, it is challenging for S-KPT to identify the left ukulele as the source of the sound first.
However, our M-KPT module aids in recognizing that the left ukulele is active in the first two segments due to its large movements, as evidenced by the motion-activation matrix $\bm{m}$ (Eq.~\ref{eq:motion_act_matrix}).
Afterward, the patches activated by the S-KPT and M-KPT are processed using the Q-KPT to retain only those patches highly relevant to the question.
Some irrelevant visual patches, such as the floor and the leg of the person on the left, are removed at this stage.
Through the collaboration of three modules, our model can focus on the most informative clues, ensuring accurate answers.

\noindent\textbf{Visualization on the impacts of the $\mathcal{A}_t^m$ and $\mathcal{A}_t^s$.}
Quantitative results in Table~\ref{tab:ablation_ms_adj} demonstrate that our model, trained with motion- and sound-driven adjacency matrices, outperforms the model directly adopting the vanilla feature similarity as the adjacency matrix.
In Fig.~\ref{fig:ablation_visual}, we visualize the learned adjacency matrices of both models.
For each visual patch (node in the graph), we sum weights from all adjacent patches via $\mathcal{A}_t^m$/$\mathcal{A}_t^s$.
A larger ``$\bullet$'' indicates that the visual patch receives more attention (larger weights) in graph learning.
The model trained with motion guidance (w/ $\mathcal{A}_t^m$) accurately highlights and tracks the key patches related to the two sounding instruments; otherwise (w/o $\mathcal{A}_t^m$), it emphasizes many irrelevant patches, such as the window and wall.
Similarly, the model trained with audio guidance (w/ $\mathcal{A}_t^s$) pays more attention to the left guitar emitting louder sound; otherwise (w/o $\mathcal{A}_t^s$), the model focuses on the middle guzheng, resulting in an incorrect prediction.

\section{Conclusion}
We propose a Patch-level Sounding Object Tracking (PSOT) method for the challenging AVQA task.
By exploring patch-wise motion information across adjacent frames, we design a motion-driven Key Patch Tracking (KPT) module to track salient visual patches.
Meanwhile, we propose a sound-driven KPT module to track sounding patches by exploring audio-visual correspondence.
After identifying salient and sounding patches, we further utilize a question-driven KPT module to retain tracked patches highly relevant to the question.
All these modules are implemented using graph neural networks, with audio-visual-question features automatically updated in graph learning.
Extensive quantitative and qualitative results validate the effectiveness of our method.

\section*{Acknowledgements}
We sincerely thank the anonymous reviewers for their invaluable comments and insightful suggestions. We also extend our gratitude to Dr. Hui Wang for his helpful discussions.
This work was supported by the National Natural Science Foundation of China (62272144,72188101, and 62020106007), the Major Project of Anhui Province (2408085J040, 202203a05020011), and the Fundamental Research Funds for the Central Universities (JZ2024HGTG0309, JZ2024AHST0337, and JZ2023YQTD0072).

\bibliography{aaai25}

\newpage
\appendix

\section{Supplementary Material}

\subsection{A. Additional Ablation Studies}

We provide more ablation studies on the core modules in the proposed method. Experiments in this section are conducted on the MUSIC-AVQA~\cite{li2022learning} dataset.

\noindent\textbf{A.1. Ablation study on the execution manner of M-KPT and S-KPT modules.}
In our main paper, the proposed M-KPI and S-KPI modules are executed in parallel.
We further explore two variants where these two modules are executed in series: a) The M-KPT module is first applied to identify and track salient visual patches, obtaining the motion-enhanced features. Based on the M-KPT module, the S-KPT is then performed to further highlight sounding patches.
b) The S-KPT module is first executed, followed by the M-KPT module.
As shown in Table~\ref{tab:execution}, we empirically find that executing these two modules in parallel (c) results in a slightly higher average performance.
We speculate that the model is able to achieve a better balance in the identification of salient and sounding patches guided by parallel motion and audio guidance.

\begin{table}[ht]
  \scriptsize
  \centering
  \setlength\tabcolsep{8pt}
  \small
  {
   \begin{threeparttable}
   \resizebox{1\columnwidth}{!}{
        \begin{tabular}{c | c|ccc|c}
            \Xhline{1.2pt}
            Id & Execution Manner &A-QA & V-QA &AV-QA  & \textit{Avg.} \\
            \hline
            a &M-KPT $\rightarrow$ S-KPT &{79.89} &86.84 &73.71 & 78.20 \\
            b &S-KPT $\rightarrow$ M-KPT &79.40 &{87.16} &73.26 &78.12 \\
            \hline
            c &M-KPT $||$ S-KPT & 79.08 & 87.12 & {74.07} & {78.42} \\
            \Xhline{1.2pt}    
        \end{tabular}}
            
        \end{threeparttable}
    }
    \caption{ \textbf{Comparison between serial and parallel execution manners of M-KPT and S-KPT modules.} `$\rightarrow$' denotes that data flow is serial; `$||$' denotes the parallel manner.}
\label{tab:execution}
\end{table}

\noindent\textbf{A.2. Impact of the parameter $\lambda$ used in M-KPT module.}
The parameter $\lambda$ is used to balance the local and global motion matrix when constructing $\mathbf{m}_t$ (Eq. 2 in our main paper). 
To explore its impact, we perform a parameter study.
As shown in Table~\ref{tab:ablation_lambda}, the model has slightly worse performance when only using the local (\ie, $\lambda=0$
) or global (\ie, $\lambda=1$) motion information.
The optimal performance is achieved when $\lambda=0.2$, which is used as the default configuration in our experiments.
These results also indicate that it is beneficial to simultaneously consider the motion guidance from local and global temporal segments.

\begin{table}[h]
  \scriptsize
  \centering
  \setlength\tabcolsep{10pt}
  \small
  {
   \begin{threeparttable}
   \resizebox{0.99\columnwidth}{!}{
        \begin{tabular}{c | c|ccc|c}
            \Xhline{1.2pt}
            \multirow{2}{*}{Id} & \multirow{2}{*}{$\lambda$ } & \multicolumn{4}{c}{Acc. (\%)} \\
            \cline{3-6}
            ~ &  ~ &\multirow{1}{*}{A-QA} & \multirow{1}{*}{V-QA} & \multirow{1}{*}{AV-QA}  & \multirow{1}{*}{\textit{Avg.}} \\
            \hline
            a & $\lambda=0 \;\;$ &79.27  &87.53  &73.60  &78.30 \\
            b & $\bm{\lambda}=\bm{0.2}$ &79.08  &87.12  &{74.07}  &\textbf{78.42} \\
            c & $\lambda=0.4$ & {79.39} &87.78 &73.64 & 78.40 \\
            d & $\lambda=0.8$ &79.45  &87.45  &73.60  &78.31 \\
            e & $\lambda=1.0$ &79.02  &87.65  &73.50  &78.23 \\
            \Xhline{1.2pt}
        \end{tabular}
        }
    \end{threeparttable}}

    \caption{
        \textbf{Impact of the hyper-parameter $\lambda$ used in M-KPT module.}
    }
\label{tab:ablation_lambda}
\end{table}

\noindent\textbf{A.3. Different manners for motion computation in M-KPT module.}
In our main paper, we quantify patch-wise motion intensity using the feature similarity of adjacent frames.
Here, we compare this strategy with another variant that utilizes the traditional optical flow method, \ie, Farneback~\cite{farneback2003two} for motion estimation.
As shown in Table~\ref{tab:optial_flow}, our feature similarity-based strategy shows a slight performance advantage. The results also verify the effectiveness and robustness of motion cues across different measurements. However, optical flow calculations are time-intensive, while using cosine feature similarity can speed up computation.

\begin{table}[h]
  \scriptsize
  \centering
  \setlength\tabcolsep{7pt}
  \small
  {
   \begin{threeparttable}
   \resizebox{0.99\columnwidth}{!}{
        \begin{tabular}{c |ccc|c}
            \Xhline{1.2pt}
             Motion Measurement &\multirow{1}{*}{A-QA} & \multirow{1}{*}{V-QA} & \multirow{1}{*}{AV-QA}  & \multirow{1}{*}{\textit{Avg.}} \\
            \hline
            optical flow & 79.02 & \textbf{87.54} & 73.77 & 78.31 \\
            \hline
            feature similarity & \textbf{79.08} & 87.12 & \textbf{74.07} & \textbf{78.42 }\\
            \Xhline{1.2pt}
        \end{tabular}
        }
    \end{threeparttable}}

    \caption{
        \textbf{Different manners for motion computation.}
    }
\label{tab:optial_flow}
\end{table}

\noindent\textbf{A.4. Impact of reservation ratio $r$ in Q-KPT module.}
$r$ is a hyperparameter used to control the number of reserved visual patches relevant to the question.
The larger the ratio $r$, the greater the number of patches reserved.
As shown in Table~\ref{tab:ablation_sparse}, using all visual patches in Q-KPT (\ie, $r$=100\%) results in the worst performance, indicating the necessity of dropping patches that are not highly relevant to the question.
The model achieves the best performance at $r$=80\%, which is set as the default configuration in our experiments.
The model's performance remains stable when $r$ gradually decreases to 40\%.
Interestingly, our model can still achieve 77.83\% accuracy utilizing only 20\% of the visual patches.
This performance even surpasses that of the model utilizing 100\% visual patches.
These results validate the superiority of the patch reservation mechanism in our Q-KPT module.

\begin{table}[ht]
  \scriptsize
  \centering
  \setlength\tabcolsep{11pt}
  \small
  {
   \begin{threeparttable}
        \begin{tabular}{ c|ccc|c}
            \Xhline{1.2pt}
             $r$ (\%)  & {A-QA} & {V-QA} & {AV-QA}  & \textit{Avg.} \\
            \hline
            $100\%$ & 78.52 & 86.09 & 72.73 & 77.30 \\
            $80\%$ &\textbf{79.08}  &87.12  &\textbf{74.07}  &\textbf{78.42} \\
            $60\%$ &78.96  &87.61  &73.87  &78.31 \\
            $40\%$ &78.90  &\textbf{87.78} &73.58  &78.28 \\
            $20\%$ &77.72  &86.62  &73.69  &77.83 \\
            
            \Xhline{1.2pt}
        \end{tabular}
    \end{threeparttable}}
    \caption{
        \textbf{Impact of reservation ratio $r$ in Q-KPT module}. 
    }
\label{tab:ablation_sparse}
\end{table}

\begin{figure}[ht]
    \centering
    \includegraphics[width=.99\columnwidth]{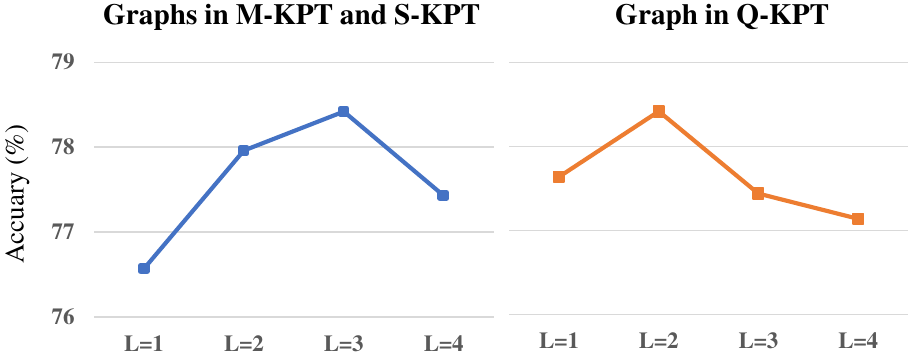}
    \caption{
    \textbf{Ablation study on the number of graph layers.}}
    \label{layer_graph}
\end{figure}

\noindent\textbf{A.5. Ablation study on the number of graph layers.}
The main modules of our proposed framework, namely M-KPT, S-KPI, and Q-KPI, are implemented using graph neural networks.
In our experiments, we empirically set the number of graph layers to 3, 3, and 2, respectively, for M-KPT, S-KPT, and Q-KPT.
Here, we provide more experimental details.
As shown in Fig.~\ref{layer_graph}, for both M-KPT and S-KPT modules, the model performance is consistently improved when the number of graph layers gradually increases from 1 to 3.
However, when continuing to use deeper graph layers, the model performance exhibits an obvious decrease.
For the Q-KPT module, the peak of the model's performance is achieved when the number of graph layers is set to 2.
Accordingly, we set the optimal values for the graph construction.

\begin{table*}[!t]
  \scriptsize
  \centering
  \setlength\tabcolsep{8pt}
  \small
  \scalebox{1}{
       \begin{tabular}{c|cccccccc|c}
       \Xhline{1.2pt}
       Method &Which &From & Happening &Where & Why &Bef.Next &When &Used-For & \textit{Avg.} \\
       \hline
       LADNet~\cite{li2019learnable} &84.2 &89.0 &79.1 &81.4 &\textbf{68.6} &82.0 &52.4 &76.5 &84.1\\
       PSCA~\cite{li2019beyond} &89.0 &91.1 &83.2 &81.7 &61.6 &82.0 &52.4 &76.5 &87.4 \\
       HGA~\cite{jiang2020reasoning} &88.6 &92.2 &83.8 &82.6 &61.6 &78.0 &52.4 &\textbf{82.4} &87.7 \\
       ACRTF~\cite{zhang2020action} &88.5&91.7&83.9&84.9&50.0&82.0 &57.1 &64.7&87.8 \\
       HCRN~\cite{le2021hierarchical} &89.8 &92.8 &86.0 &84.4 &57.0 &80.0 &52.4 &\textbf{82.4} &89.0 \\
       PSTP~\cite{li2023progressive} &-- &--  &--  &--  &--  &-- &-- &-- &90.2 \\
       \hline
       \textbf{PSOT (ours)} &\textbf{91.5} &\textbf{93.8} & \textbf{86.7} &\textbf{88.1} &\underline{65.1} &\textbf{91.8} &\textbf{57.4} &\underline{79.2} &\textbf{91.3} \\
       \Xhline{1.2pt}
       \end{tabular}
    }
    
    \caption{
        \textbf{Comparison with prior works on the AVQA~\cite{yang2022avqa} dataset.}    }
    \label{tab:avqa}
\end{table*}

\noindent\textbf{A.6. Ablation study on the MMA module.}
We propose the Multimodal Message Aggregation (MMA) module to integrate clues from visual, audio, and question modalities for final answer prediction (Sec. 3.4 in main paper).
This module utilizes segment-level audio features $\bm{\overline{a}} \in \mathbb{R}^{T \times d}$, patch-level visual features $\bm{\hat{v}} \in \mathbb{R}^{T \times N^2 \times d}$, and segment-level visual features $\bm{\overline{v}} \in \mathbb{R}^{T \times d}$ to interact with the question features.
We perform ablation experiments to explore the impacts of these audio and visual features.
As shown in Table~\ref{tab:fusion_module_ablation}, compared to the full MMA module (a), the model performance significantly decreases when the audio or visual features $\bm{\overline{a}}$, $\bm{\hat{v}}$, or $\bm{\overline{v}}$ (b/c/d) are removed, respectively.
These results indicate that it is beneficial to consider audio-visual features at different levels for the multimodal AVQA task.

\subsection{B. Evaluation on additional AVQA~\cite{yang2022avqa} Dataset}
In our main paper, we have compared our method with prior works on the widely-used MUSIC-AVQA~\cite{li2022learning} dataset.
To examine the robustness of our method, we further evaluate our method on the additional AVQA~\cite{yang2022avqa} dataset, which contains over 57,000 videos related to real-life audio-visual scenarios.
Table~\ref{tab:avqa} shows the performance comparison between our method with prior works.
Our method surpasses the previous state-of-the-art method PSTP~\cite{li2023progressive} by 1.1\% in terms of the average performance.
These results further verify the effectiveness and superiority of the proposed method.

\begin{table}[t]
  \scriptsize
  \centering
  \setlength\tabcolsep{10pt}
  \small
  {
   \begin{threeparttable}
   \resizebox{1\columnwidth}{!}{
        \begin{tabular}{c | c|ccc|c}
            \Xhline{1.2pt}
            Id & Method &A-QA & V-QA &AV-QA  & \textit{Avg.} \\
            \hline
            a & full &\textbf{79.08} &\textbf{87.12} &\textbf{74.07} &\textbf{78.42}\\
            b & w/o $\bm{\overline{a}}$ &74.74 &\underline{85.18} &\underline{71.18} &75.52 \\
            c & w/o $\bm{\overline{v}}$ &\underline{78.96} &84.02 &71.04 &\underline{75.88}  \\
            d & w/o $\bm{\hat{v}}$ &78.46 &83.20 &70.81 &75.45 \\
            \Xhline{1.2pt}
        \end{tabular}}
        \end{threeparttable}
        
    }
    \caption{ \textbf{Ablation study on MMA module.} Top-2 results are highlighted in bold and \underline{underlined}, respectively.}
    \label{tab:fusion_module_ablation}
\end{table}

\subsection{C. More Discussions on Motion Guidance}
Given the diversity of audio-visual scenes in the wild, the motion of sounding objects can vary over time. Therefore, our proposed PSOT method must account not only for situations with salient motion cues but also for those where the motion intensity is less pronounced.  However, we found that the motion cues are generally beneficial, as evidenced by our experiments (Table~\ref{tab:main_module_abltion}\&~\ref{tab:ablation_ms_adj}).
We provide more discussions here.

\begin{itemize}
    \item In many cases, visual patches with large movements (salient patches) correspond to/relate closely to sounding objects and questions. Even if the visual patches do not exactly match the sounding objects (e.g., a hand near a sounding guitar), they often provide context that directs the model’s attention to nearby patches.

    \item Even under cases where salient patches with large motion may not match the sound, our method incorporates M-KPT, S-KPT, and Q-KPT modules, leveraging extra cooperation from sound and question cues as additional guidance to refine/update key patches. We measure motion magnitude from the motion intensity matrix (Eq.~\ref{eq:motion_act_matrix}), dividing test videos into low (0-30\%), moderate (30-70\%), and high (70-100\%) motion ranges. Table~\ref{tab:motion_results} suggests that our model maintains strong performance even at lower motion levels, due to cooperative mechanism.

    \begin{table}[t]
      \scriptsize
      \centering
      \setlength\tabcolsep{8pt}
      \small
      {
       \begin{threeparttable}
       \resizebox{0.99\columnwidth}{!}{
            \begin{tabular}{ c|ccc|c}
                \Xhline{1.2pt}
                 Magnitude (\%)  & {A-QA} & {V-QA} & {AV-QA}  & \textit{Avg.} \\
                \hline
            0-30                & 79.25         & 87.06         & 73.87          & 78.30        \\ \hline
            30-70               & 78.68         & 87.25         & 74.08          & 78.44        \\ \hline
            70-100              & 78.76         & 87.34         & 74.15          & 78.51       \\
            \Xhline{1.2pt}
            \end{tabular}
            }
        \end{threeparttable}}
    \caption{\textbf{Model performance across various motion magnitudes.}}
    \label{tab:motion_results}
    \end{table}

    \item Motion cues are particularly valuable for distinguishing between sounding objects with similar appearances. As shown in Fig.1(a), while there is a clear guitar sound in the last video segment, identifying which guitar, or if both guitars are producing sound, remains challenging. By recognizing that both guitars’ visual patches exhibit large movements, our model can infer that both guitars are producing sound, thereby facilitating accurate question answering. This case is often ignored by previous methods.
    
    \item The AVQA task also handles cases which questions relate only to visual signals, making visual motion cues especially valuable.
\end{itemize}

\end{document}